\documentclass[12pt,preprint]{aastex}

\slugcomment{Accepted by ApJL}

\shorttitle{A New Method to Study the Origin of the EGB}
\shortauthors{Zhou et al.}

\begin{document}


\title{A New Method to Study the Origin of the EGB \\and the First Application on AT20G}

%
\author{
Ming Zhou\altaffilmark{1,2,3},
Jiancheng Wang\altaffilmark{1,2},
Xiaoyan Gao\altaffilmark{1,2,3}
}
\altaffiltext{1} {National Astronomical Observatories, Yunnan Observatory, Chinese Academy
of Sciences,  Kunming 650011, China}
\altaffiltext{2} {Key Laboratory for the Structure and Evolution of Celestial Objects,
Chinese Academy of Sciences,  Kunming 650011, China}
\altaffiltext{3} {Graduate School, Chinese Academy of Sciences, Beijing, P.R. China}


\email{mzhou@ynao.ac.cn}



\begin{abstract}
In this letter, we introduce a new method of image stacking to directly study the undetected
but possible $\gamma$-ray point sources. Applying the method to
the Australia Telescope 20 GHz Survey (AT20G) sources which have not been
detected by Large Area Telescope (LAT) on {\it Fermi Gamma-ray Space Telescope} ({\it Fermi})
, we find that the sources contribute (10.5$\pm$1.1)\,\% and (4.3$\pm$0.9)\,\%  of the
extragalactic gamma-ray background (EGB)
and have a very soft spectrum with the photon indexes of 3.09$\pm$0.23 and 2.61$\pm$0.26, 
in the 1--3 and 3--300\,GeV energy ranges. In the 0.1--1\,GeV range, they probably contribute more large
 faction to the EGB, but it is not quite sure. 
It maybe not appropriate to assume that the undetected
sources have the similar property to the detected sources.
\end{abstract}



\section{Introduction}
The EGB was first detected by the SAS-2 mission \citep{fichtel75}, and its  spectrum
was measured with good accuracy by {\it Fermi} \citep[also called isotropic diffuse background,][]{abdo10a}.
It has been found to be consistent with a featureless power law with
a photon index of $\sim$2.4 in the 0.2--100\,GeV energy range and an integrated flux (E$\geq$100\,MeV)
of 1.03$\times10^{-5}$\,ph cm$^{-2}$ s$^{-1}$ sr$^{-1}$.

The origin of the EGB  is one of the fundamental unsolved problems in astrophysics, and it has
been a subject of study for a long time \citep[see][for a review]{k08}. The EGB could originate
from either truly diffuse processes or from unresolved point sources. Truly diffuse emission can
arise from numerous processes such as the annihilation of dark matter \citep{ahn07,c10,b10},
particle acceleration by intergalactic shocks produced during large scale structure
formation \citep{gb03} etc.

Blazars (including BL Lac objects, flat spectrum radio quasars, or unidentiffed flat spectrum radio sources)
represent the most numerous population detected by the Energetic Gamma Ray Experiment Telescope
(EGRET) on Compton Gamma Ray Observatory \citep{h99} and {\it Fermi} \citep{abdo10d}.
Therefore, the blazars which have not been detected by the EGRET or
LAT are the most likely candidates for the origin of the bulk of the EGB emission.
Many authors have  studied the luminosity function of blazars and showed that the contribution of blazars to
the EGRET EGB could be in the range from 20\,\% to 100\,\% \citep{s96,n06,d07,c08,kn08,i09}.
Nevertheless, starburst galaxy and non-blazar radio loud active galactic nuclei can also contribute
a fairly large fraction of the EGB \citep{t07,bs09,b09}.

Recently, \citet{abdo10b} built a source count distribution at GeV energy and yielded that
point sources which had not been detected by the LAT can contribute 23\,\% of the
EGB. At the fluxes currently reached by the LAT, they ruled out the hypothesis that point-like
sources (i.e.\,blazars) produce a large fraction of the EGB.

However, if the property of undetected sources is not  similar to the detected sources,
these conclusions maybe not correct.
Therefore, we apply an image stacking method to directly study the undetected point sources.
For a sample of possible $\gamma$-ray point sources which have not been detected by the
{\it Fermi} due to their faint fluxes or soft spectra \citep{abdo10c}, we can stack a large number of them to
improve the statistics \citep{ando10}. If their fluxes are not too faint, we can derive their
mean flux and photon index by Maximum Likelihood (ML) method.

\section{Sample}
The AT20G\footnote{It is available online through Vizier (http://vizier.u-strasbg.fr)}
\citep{murphy10} is the largest catalog of high frequency  radio sources and contains
5890 sources with the flux at 20\,GHz exceeding 40 mJy in the whole sky south of declination 0$^\circ$.
In the south sky, about 60\,\% (230 sources) of {\it Fermi} 1--year LAT AGN Catalogue (1LAC) sources
are associated with the AT20G sources \citep{G10a}. Through studying the correlation between the
$\gamma$-ray and radio flux density of AT20G sources, \citet{G10b} yielded that AT20G sources not detected by the LAT 
can contribute 17\,\% of the EGB. Therefore, we apply our methods to study them firstly.

We exclude the sources identified as Galactic H$_{\rm II}$ regions, Galactic Planetary Nebulas and
parts of the Magellanic Clouds in AT20G, then the majority of  sources (5808) we obtain  are
quasi-stellar objects \citep[see][but the optical properties of these objects have not  been
published]{murphy10}.  In order to minimize the influence of strong sources, we only use the
sources that  are at least $2^\circ$ away from the nearest  First {\it Fermi}-LAT catalog (1FGL)
 source and locate at high Galactic latitudes, $|b|>15^\circ$. Finally, we obtain
 2900 sources to analyze their contribution to the EGB.

 \section{Method}\label{method}
The photons\footnote{http://fermi.gsfc.nasa.gov/cgi-bin/ssc/LAT/LATDataQuery.cgi} we used in
our analysis are same as that used by \citet{abdo10c} to construct the 1FGL,
but ours in the  1--300\,GeV energy range have small 68\,\% containment radius
(better than 1$^\circ$) and little source confusion \citep[see][]{at09,abdo09}. In this procedure, the  tools
of {\it gtselect} and {\it gtmktime}\footnote{These and other tools we used in next are accessible at \\
http://fermi.gsfc.nasa.gov/ssc/data/analysis/scitools/overview.html} are used.

For stacking the images of all sources, we collect all photons that are at most  $1^\circ$ away from any source
of our sample and then record their energies ($E_i$, in units of GeV) and angular
distances ($\theta_i$, in units of deg) between the photon and the source. The overlaping of sources
have little influence on our method because these sources are very faint and can be regard as parts of
diffuse background source,  especially in stacked image.

After that, we apply a ML method to derive the  flux and photon index of the
stacked point source. For simplicity, in our model there are only two sources, e.g. the
diffuse background source and stacked point source. We assume that they all have  power-law spectra,
and the photon indexes are  $\gamma_1$ and $\gamma_2$, respectively. The fluxes density are
$f_1$ (in units of [ph cm$^{-2}$ s$^{-1}$ GeV$^{-1}$ deg$^{-2}$]) and
$f_2$ (in units of [ph cm$^{-2}$ s$^{-1}$ GeV$^{-1}$]), respectively. The emission can be described by
\begin{equation}
 \frac{dN(\theta,E)}{2\pi\theta d\theta dE}=[ f_1(\frac{E}{1\,\rm GeV})^{-\gamma_1}+
f_2(\frac{E}{1\,\rm GeV})^{-\gamma_2}PSF(\theta,E)] exposure (E),
\end{equation}
where $dN(\theta,E)$ is the photon number in the ranges
of ($\theta$--$\theta\!+\!d\theta$) and ($E$--$E\!+\!dE$), PSF is the point spread function (in the units of
deg$^{-2}$), {\it exposure} is the  integral of effective area over time (in units of [cm$^2$ s]).
 PSF and {\it exposure} are derived from the tool of {\it gtpsf}.  The emission must meet the relationship of
\begin{equation}
\int_{0}^{1}\int_{E_1}^{E_2} [ f_1(\frac{E}{1\,\rm GeV})^{-\gamma_1}+
f_2(\frac{E}{1\,\rm GeV})^{-\gamma_2}PSF(\theta,E)] exposure (E) 2\pi\theta d\theta dE=N_0,
\end{equation}
where $N_0$ is the total number of photons in the $E_1$--$E_2$ energy range. Therefore, there are
only three free parameters. In the practical calculation, we use $\gamma_1$, $\gamma_2$
and $M$. $M$ is the number of photons contributed by the stacked source. Then $f_1$ and $f_2$ can be 
described by $\gamma_1$, $\gamma_2$ and $M$.

The probability for a photon with ($\theta_i$,$E_i$) is
\begin{equation}
P_i=\frac{exposure (E_i)2\pi\theta_i}{N_0}[f_1(\frac{E_i}{1\,\rm GeV})^{-\gamma_1}
+f_2(\frac{E_i}{1\,\rm GeV})^{-\gamma_2}PSF(\theta_i,E_i) ].
\end{equation}
The likelihood is the probability of the observed data for a specific model. For our case, it is defined as:
\begin{equation}
L=\prod_{i=1}^{N_0} P_i.
\end{equation}
The logarithm of the likelihood is much conveniently calculated
\begin{equation}
\ln{L}=\sum_{i=1}^{N_0}\ln[f_1(\frac{E_i}{1\,\rm GeV})^{-\gamma_1}
+f_2(\frac{E_i}{1\,\rm GeV})^{-\gamma_2}PSF(\theta_i,E_i)]
+\sum_{i=1}^{N_0} \ln \frac{exposure (E_i)2\pi\theta_i}{N_0}.
\end{equation}
Because the last term is model independent, it is not useful for ML method. Neglecting the last term, we get
\begin{equation}
\ln{L}=\sum_{i=1}^{N_0}\ln[f_1(\frac{E_i}{1\,\rm GeV})^{-\gamma_1}\\
+f_2(\frac{E_i}{1\,\rm GeV})^{-\gamma_2}PSF(\theta_i,E_i)].
\end{equation}
We maximize numerically L to obtain the most probable parameters
($f_1$, $f_2$ and $\gamma_1$, $\gamma_2$) of these sources.

We use the likelihood ratio to test the hypothesis. The point source ``test statistic'' (TS)
is defined as
\begin{equation}
TS=-2(\ln{L_0}-\ln{L_1}),
\end{equation}
where $L_0$ and $L_1$ is the likelihood without and with point source. The detected significance of
a point source is approximately $\sqrt{TS}\sigma$ \citep[see][]{m96}.

\section{Result \& Discussion}
The stacked source is estimated to has a photon index of 2.81 and integrated flux of
1.07$\times10^{-7}$\,ph cm$^{-2}$ s$^{-1}$. 
The TS is 129, corresponding to a significance of $\sim$11$\sigma$.
The mean flux of these sources is 3.69$\times10^{-11}$\,ph cm$^{-2}$ s$^{-1}$,
it is fainter than the faintest 1FGL source by a factor of 10.
 It can contribute 8.4\,\% of the EGB in the 1--300\,GeV energy range.
  We also apply our method to a subsample of flat spectrum radio sources
  (i.e. $\alpha_{\rm (5-20GHz)}<$0.5, with  $F_{\nu}\propto\nu^{-\alpha}$, 1780 sources).
 Its photon index is 2.79, only slightly harder than the former. Its mean flux is
 3.79$\times10^{-11}$\,ph cm$^{-2}$ s$^{-1}$, and the TS is 88. This subsample has not
 distinct characteristic from the other sources in $\gamma$-ray energy range.
%

In order to test a more complicated spectral shape of the stacked source, we analyze the spectrum
in the 1--10\,GeV energy range . We expect that the spectrum would be harder in this energy range.
However, the estimated photon index is 3.01. Therefore, we analyze the spectrum in 2--10\,GeV,
3--10\,GeV energy range, respectively. The results are summarized in Table 1. We find
that the spectrum is very soft in 1--3\,GeV energy range and becomes harder above 3\,GeV. 
It is indicated that two types of sources exist,  in which one with softer and another with harder spectrum
in GeV range. The former will dominate in lower energies, and latter in higher energies.
Therefore, the spectrum of stacked source shows very soft in the 1--3\,GeV energy range.
We will study this further if the optical properties of these objects can be obtained.

Finally we obtain the properties of the spectrum as follows. In the 1--3\,GeV, the photon index is 3.12, 
the mean integrated flux is 3.89
$\times10^{-11}$\,ph cm$^{-2}$ s$^{-1}$, and the TS is 92. Obviously the flux is larger than
that in the 1--300\,GeV energy range. 
It could be caused by that the spectrum in the 1--300\,GeV is not well fitted with a single power-law.
In the 3--300\,GeV, the photon index is 2.66, the mean integrated flux is 3.72
$\times10^{-12}$\,ph cm$^{-2}$ s$^{-1}$, and the TS is 38.

An decrease in $\ln{L}$ of 0.5 from its maximum value corresponds to the 68\% confidence (1 $\sigma$)
region for each parameter \citep[see][]{m96}. We use this variance to estimate the error of each parameter.
In three parameters ($\gamma_1$, $\gamma_2$ and $M$), we take two ones to be the values with maximal likelihood,
and allow third one to change around its best value, we then test the deviation of $\ln{L}$ from
its maximum value shown in figure 1. The 1 $\sigma$ errors of $\gamma_2$ are 0.25 and 0.22, the ones of 
$M$ are 270 and 53, in 1--3 and 3--300\,GeV respectively. The 1 $\sigma$ relative errors
of fluxes are 10.5 \% and 17.3 \%, in 1--3 and 3--300\,GeV.

In order to verify the effectiveness and accuracy of our method, we do the Monte Carlo (MC)
simulations using the tool {\it gtobssim}. The simulating time is 26\,Ms, equaling to the time of real data we used. 
We simulate the Galactic and isotropic diffuse backgrounds using the models
(e.~g. gll\_iem\_v02.fit, isotropic\_iem\_v02.txt) recommended by the LAT team, in which 3913 sources 
are generated each time, but  only 2900 sources isotropically distribute on the sky with $|b|>15^\circ$.
We complete  one thousand MC simulations in 1--3 and 3--300\,GeV energy range using the obtained parameters.
The diffuse source is simulated only once due to long run time, but its effect on the result is not remarkable
 because the source is random distribution and its photons are various.

The distributions of the photon index, flux and TS for different energy ranges are shown in figure 2. They are compatible
with Gaussian distributions. In the 1--3 and 3--300 \,GeV, the mean fluxes are 4.30 and 0.364
(in the unit of [$10^{-11}$\,ph cm$^{-2}$ s$^{-1}$]), their relative errors are 
10.3\,\% and 20.7\,\%; the photon indexes are 3.15 and 2.71 with the errors of 0.23 and 0.26.
The errors estimated here are similar to that found in the fourth paragraph.
Comparing the input parameters, we find that the systematic errors occur, especially for the flux in the
1--3\,GeV energy range. They could be caused by that diffuse background source can not
be described by a single power-law spectrum.

Because the MC method can obtain the systematic errors, we use this method to correct our results as follows: 
in the 1--3 and 3--300\,GeV, the fluxes are 3.48$\pm$0.36 and 0.380$\pm$0.080 
(in unit of [$10^{-11}$\,ph cm$^{-2}$ s$^{-1}$]),
and the photon indexes are 3.09$\pm$0.23 and 2.61$\pm$ 0.26 respectively, while
the contribution to the EGB is (10.5$\pm$1.1)\,\% and (4.3$\pm$0.9)\,\%, which are much
smaller than the result (17\,\%) of \citet{G10b}.
If the soft spectrum in 1--3\,GeV is caused by the spectral broken of some sources, the
photon index would not be extrapolated to lower energy range.  However, as long as the spectrum of
stacked source is not harder than the EGB, the contribution to the EGB in 0.1--1\,GeV
will be larger than that in 1--3\,GeV. Our result is compatible with the result (23\,\%) of \citet{abdo10b} 
because other point sources could contribute to the EGB.

In this letter, we introduce a new method of images stacking to directly study the contribution
of undetected point sources to the EGB. Our method is more direct than the methods used by many authors. 
Those methods involve the $\gamma$-ray luminosity of undetected sources which is estimated through 
the properties of a few detected sources.  They include many uncertainties and lead the result to be questionable validity.

Applying our method, we find that the undetected sources in AT20G can contribute
(10.5$\pm$1.1)\,\% and (4.3$\pm$0.9)\,\% to the EGB in the 1--3 and 3--300\,GeV
energy range respectively. Their $\gamma$-ray spectrum is very soft, implying that the emissive property is different for 
undetected and detected sources.

Applying our method to estimate the contribution of all point sources to the EGB, we need a complete 
sample of possible $\gamma$-ray point sources 
which is not easily constructed. In this letter, we only estimate the contribution of AG20G to the EGB. We will
study more samples of possible $\gamma$-ray point sources in the future.

\acknowledgments
\section*{Acknowledgments}
We thank the LAT team and AT20G team providing the data on the website.
We acknowledge the financial supports from the National Natural
Science Foundation of China 10778702, the National
Basic Research Program of China (973 Program 2009CB824800), and the
Policy Research Program of Chinese Academy of Sciences (KJCX2-YW-T24).

\clearpage

\begin{table}
\begin{center}
\caption{The results in different energy ranges.}
\begin{tabular}{ccccccc}

\hline\hline
Energy Ranges &Photon Indexes&Mean Fluxes&TS\\
(GeV)&  &($10^{-12}$ ph cm$^{-2}$ s$^{-1}$)&\\
\hline
1--300&2.81&36.9&129\\
1--10&3.02&38.4&108\\
2--10&2.95&7.49&37\\
3--10&2.64&2.84&19\\
1--3&3.12&38.7&92\\
3--300&2.66&3.72&38\\
\hline
\hline
\end{tabular}
\end{center}
\end{table}

\begin{figure}
\begin{center}
\includegraphics[height=0.9\textwidth,width=0.9\textwidth]{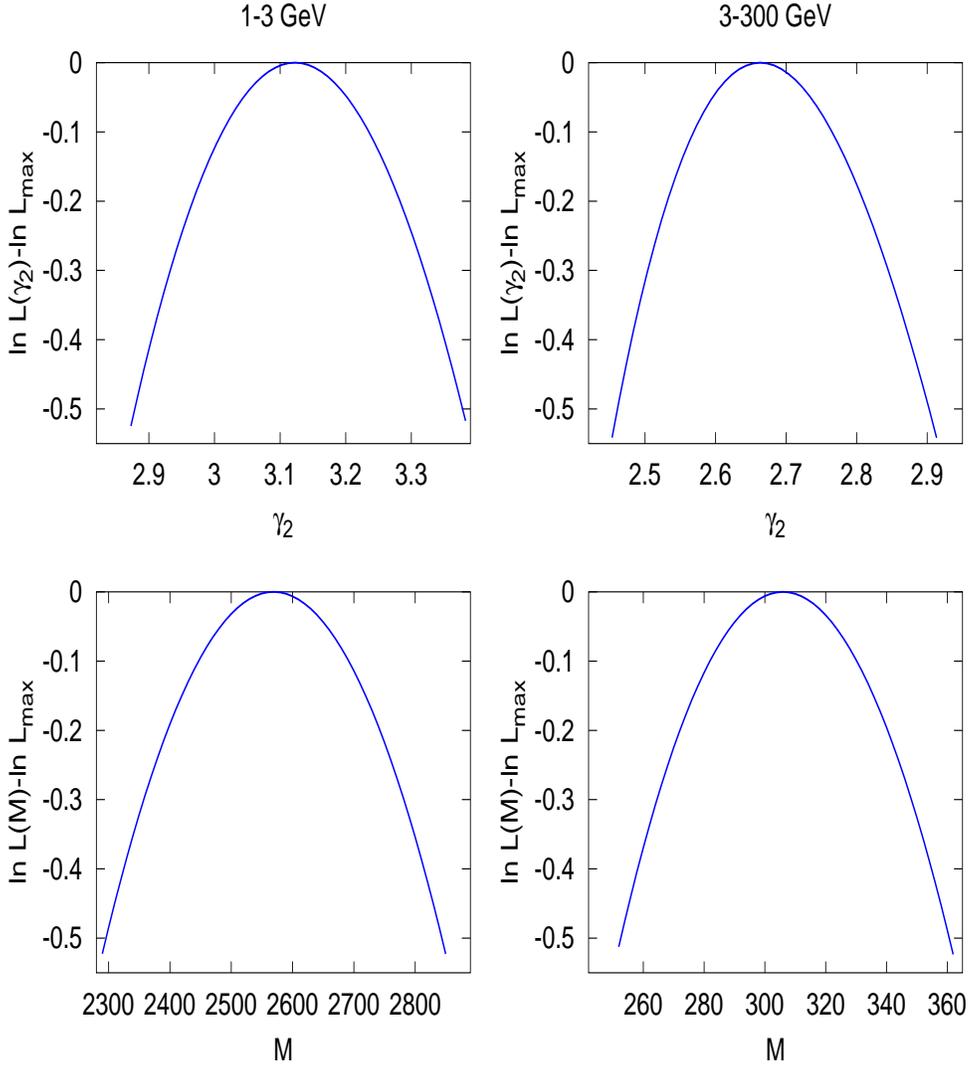}
\caption{The change of likelihood from its maximum with $\gamma_2$ (top) or
$M$ (bottom) around their best value when other parameters are fixed on their highest
likelihood value. Right and left panels are for 1--3\,GeV and 3--300\,GeV bands, respectively.
}
\end{center}
\end{figure}
\begin{figure}
\begin{center}
\includegraphics[height=0.9\textwidth,width=0.9\textwidth]{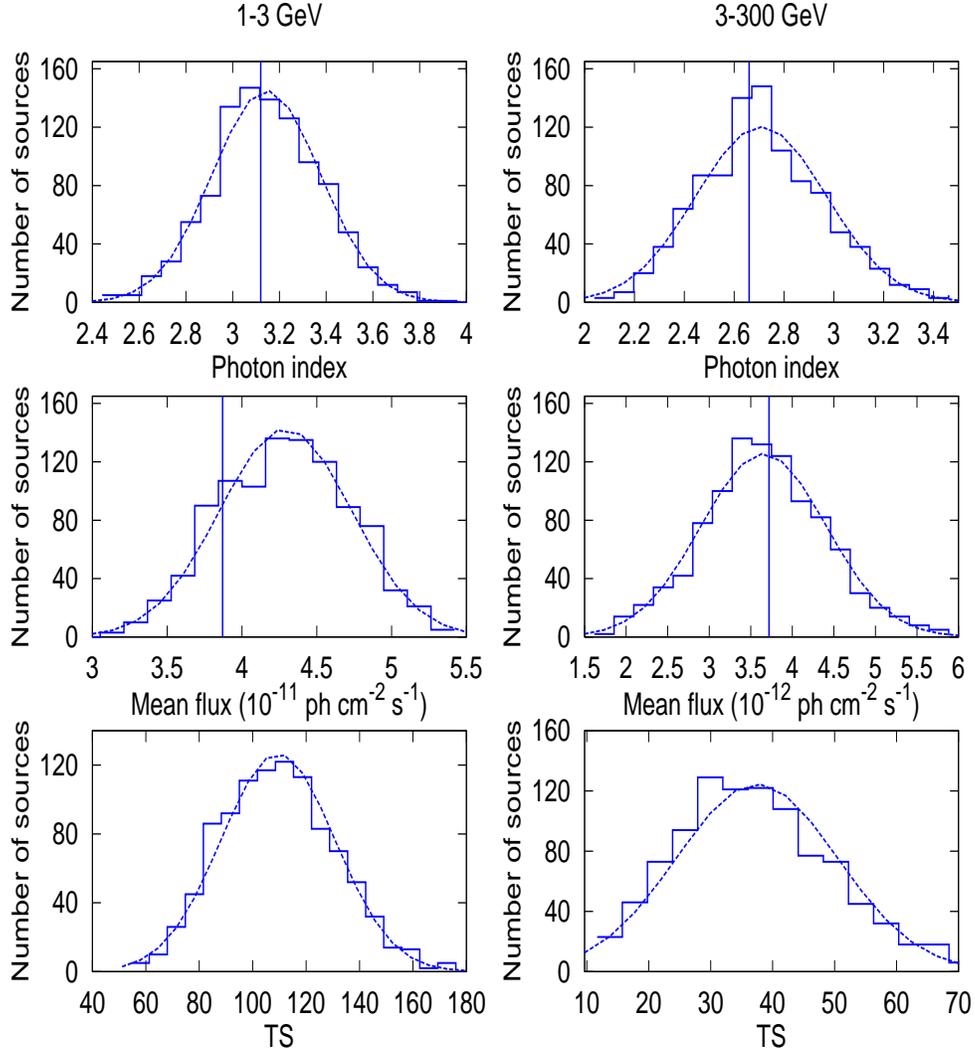}
\caption{Distributions of photon indexes (top), mean fluxes (middle) and TS (bottom).
Right and left panels are for 1--3\,GeV and 3--300\,GeV bands, respectively.
The distribution can be represented by  Gaussian functions (dashed line) with central values
$\mu$=3.15 (2.71), 4.30 (0.364) $\times10^{-11}$, 110 (37), respectively, standard deviations
$\sigma$=0.23 (0.26), 4.44 (0.755)$\times10^{-12}$,  21 (13), in the 1--3 (3--300)
\,GeV. The input photon indexes are 3.12, 2.66, fluxes are 3.87, 0.372 (in unit of
[$10^{-11}$\,ph cm$^{-2}$ s$^{-1}$]), respectively, which are indicated by vertical line}
\end{center}
\end{figure}


\end{document}